%% file: cygx2_apjlett_p1_final.tex
\documentclass{emulateapj}
\usepackage{apjfonts, epsfig}

\def\asca{{\sl ASCA }}
\def\sax{{\sl BeppoSAX }}

\def\gin{{\sl GINGA }}

\def\exo{{\sl EXOSAT }}

\def\xte{{\sl RXTE }}

\def\bbxrt{{\sl BBXRT }}

\def\chandra{{\sl Chandra }}
\def\cirx1{Cir~X-1~}
\def\cygx2{Cyg~X-2~}
\def\herx1{Her~X-1~}
\def\4u1822{4U 1822-37~}
\def\ergsec{\hbox{erg s$^{-1}$ }}
\def\ergcm{\hbox{erg cm$^{-2}$ s$^{-1}$ }}

\def\kmps{\hbox{km $s^{-1}$}}
\def\fexxiv{Fe~{\sc xxiv~}}
\def\fexxv{Fe~{\sc xxv~}}
\def\fexxvi{Fe~{\sc xxvi~}}

\def\nex{Ne~{\sc x~}}
\def\mgxii{Mg~{\sc xii~}}
\def\sixiv{Si~{\sc xiv~}}
\def\sxvi{S~{\sc xvi~}}
\def\alxii{Al~{\sc xii~}}

\def\Msun{$M_{\odot}$ }

\def\it{\sl}

\def\lapp{\ifmmode\stackrel{<}{_{\sim}}\else$\stackrel{<}{_{\sim}}$\fi}

\def\gapp{\ifmmode\stackrel{>}{_{\sim}}\else$\stackrel{>}{_{\sim}}$\fi}
\def\spose#1{\hbox to 0pt{#1\hss}}

\def\approxlt{\mathrel{\spose{\lower 3pt\hbox{$\sim$}}
        \raise 2.0pt\hbox{$<$}}}
\def\approxgt{\mathrel{\spose{\lower 3pt\hbox{$\sim$}}
        \raise 2.0pt\hbox{$>$}}}

\slugcomment{Submitted for publication to The Astrophysical Journal}

\shorttitle{ADC EMISSIONS IN CYG X-2} 

\shortauthors{SCHULZ et al.}

\begin{document}

\title{Heating in an Extended Accretion Disk Corona along the Z-Pattern in Cyg~X-2}
\author{
N. S. Schulz\altaffilmark{1},
D. P. Huenemoerder\altaffilmark{1},
L. Ji\altaffilmark{1},
M. Nowak\altaffilmark{1}.
Y. Yao\altaffilmark{2},
and
C. R. Canizares\altaffilmark{1}}
\altaffiltext{1}{Kavli Institute for Astrophysics and Space Research, Massachusetts Institute of Technology,
Cambridge, MA 02139.}
\altaffiltext{2}{CASA, University of Colorado,
Boulder, CO .}

\begin{abstract}
We observed at very high spectral resolution 
the prototype Z-source \cygx2 twice along its entire X-ray spectral variation pattern. 
In this preliminary analysis we find an extended accretion disk corona 
exhibiting Lyman $\alpha$ emissions from various H-like ions, as well as
emissions from He-like ions of Fe and Al, and Be-like ions of Fe. The brightest lines 
show a range of line broadening: H-like lines are 
very broad with Doppler velocities between 1100 and 2700 km s$^{-1}$, 
while some others are narrower
with widths of a few hundred km s$^{-1}$. 
Line diagnostics allow us for the first time to determine coronal parameters.
The line properties are consistent with a stationary,
extended up to $10^{10}$ cm, dense ($1\times10^{15}$ cm$^{-3}$), 
and hot (log$ \xi \approxgt 3; T \approxgt 10^6$ K) accretion disk corona.
We find ongoing heating of the corona along the Z-track and 
determine that heating luminosities change from
about 0.4$\times L_{Edd}$  on the 
horizontal to about 1.4$\times L_{Edd}$ on the flaring branch.
\end{abstract}


\section{Introduction}


At the end of the 1980s two classes of Low-Mass X-ray Binaries (LMXBs)
emerged based on their behavior with
respect to X-ray luminosity, spectral, and timing evolution~\citep{hasinger1989}. 
Bright LMXBs with luminosities
near the Eddington luminosity L$_{edd}$ show spectral variations in the color-color diagram (CD)
where the spectral slope and normalization evolve in a continuous pattern from a horizontal branch (HB)
to a normal branch (NB) to a flaring branch (FB) and back~\citep{schulz1989}. 
Multi-wavelength studies in the case of the prototype Z-source Cyg X-2 
have indicated that mass accretion rate seems to continuously change along the Z-track
\citep{hasinger1990, obrien2004}.
The source itself has an estimated distance of 8--11 kpc~\citep{cowley1979, smale1998}, 
its companion is an evolved
star of a mass ranging between 0.4 and 0.7 \Msun. 
The source has been known to show extensive dipping activity on the FB, which leads
to estimated angle of inclination of at least 60$^{\circ}$.
It has long been speculated that intensity and spectral changes are a consequence 
changes in geometrical and optical thickness of the accretion disk and possibly an associated
accretion disk corona (\citealt{vrtilek1988}; ADC).
It is also well established that bright LMXBs show Fe K$\alpha$ emissions
\citep{white1985, white1986, hirano1987}, which in the case of \cygx2 have been identified
with He-like \fexxv at 6.71 keV using the higher resolution spectrometer onboard \bbxrt~\citep{smale1993}.
Subsequent observations with \asca detected an additional broad emission line complex around 1 keV,
while observations with \sax introduced additional line detections 
consistent with Fe XXIV, Si XIV, and S XVI,
with fairly narrow claimed equivalent widths~\citep{smale1994, kuulkers1997}. 
The most thorough spectral analysis of \cygx2 spectra along the 
Z-spectral pattern was performed by \citet{disalvo2002}, and to date the most consistent line
detections concern the broad emission line features at 1 keV and at 6.6-6.7 keV.
Thus although the existence of line emission in \cygx2 is now fairly established,
basic line properties, i.e. precise positions, widths, and fluxes, remain undetermined.
Furthermore, since the Z-shaped spectral changes in the color-color diagram were discovered,
there has been little progress in understanding this pattern. 

\input{tab1}

\chandra observations of Z-sources to date so far have only been confirmed
in Cir~X-1 ~\citep{brandt2000, schulz2008}. In this letter
we report on the detection of broad and highly ionized emission lines in a recent long exposure
of \cygx2 with \chandra, which
allows the precise determination of line positions, line widths, and line fluxes.
We also report on the first detection of a strong line flux correlation with the Z-pattern
in \cygx2.

\section{Chandra Observations}

\cygx2 was observed with the High Energy Transmission Grating Spectrometer (HETGS, 
see \citealt{canizares2005} for a detailed description)
on 2007 August~25 (starting at 17:45:20 UT) for 70.2~ks (OBSID 8170, obs. 1) and on 
2007 August~23 (starting at 05:02:33 UT) for another 64.4~ks 
(OBSID 8599, obs. 2). The X-ray source is one of the brightest in the sky, the ASM onboard 
the \it Rossi X-Ray Timing Explorer\rm (\xte) 
recorded an average of about 0.484 Crab throughout 2005, Given \chandra 's enormous spatial
resolution, the fastest available detector readout time (~3.82 msec), as provided
in continuous clocking (CC) mode in order to mitigate pileup in the grating spectra, was chosen. 
Table~1 summarizes basic observational parameters.

\includegraphics[angle=0,width=8.4cm]{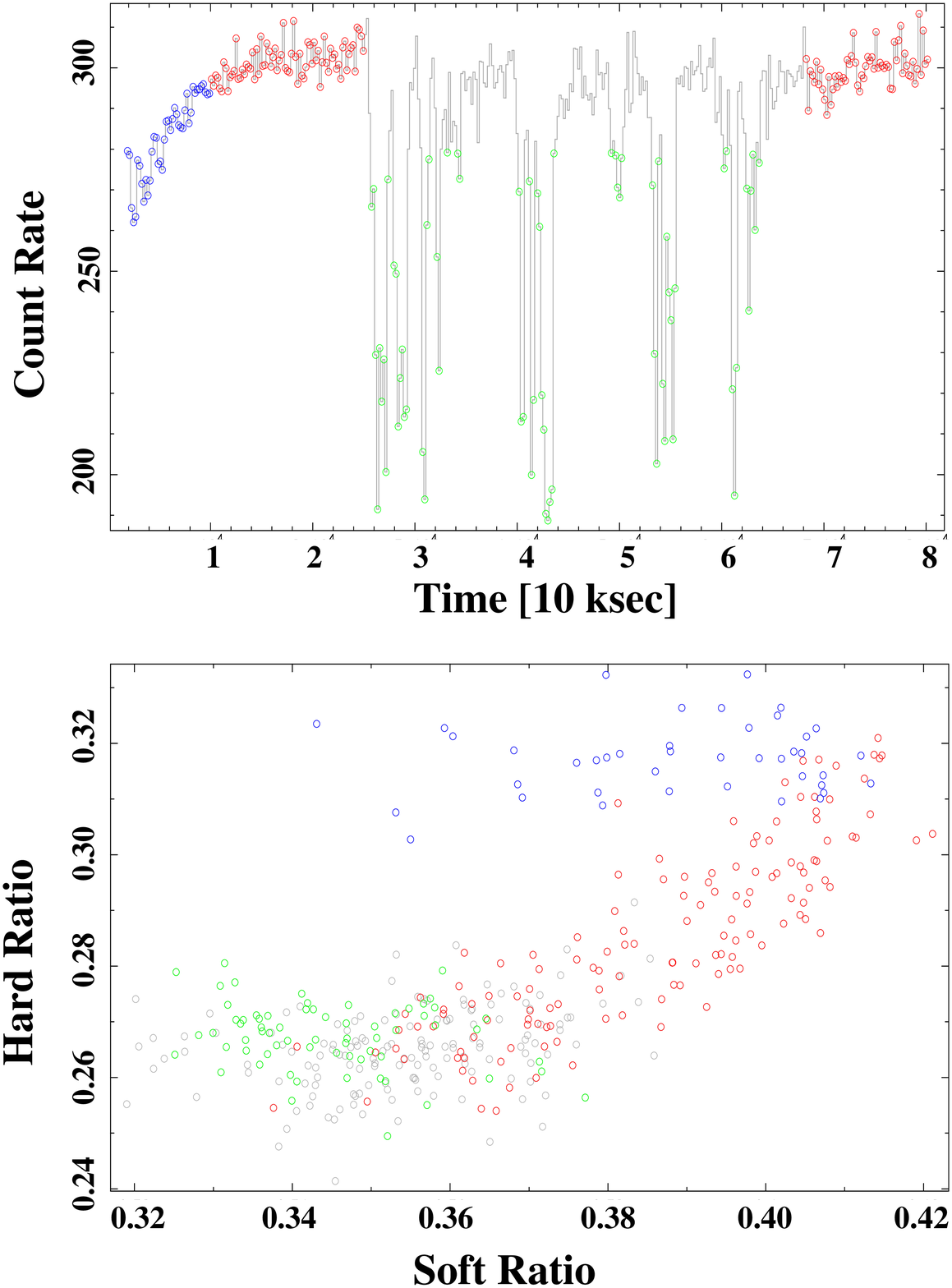}
\figcaption{{\bf Top:} The light curve of observation 1.
The colors correspond to 
the spectral branches in the color-color diagram with HB (blue), NB (red), 
and FB (green (dip) $\&$ grey (non-dip)).
{\bf Bottom:} The color-color diagram of the lightcurve above.
\label{lightcurves}} 
\vspace{0.2mm}

All data were reprocessed using CIAO4.0 and the most recent
CALDB products. The first order count rates of $\sim$ 270 cts s$^{-1}$ were quite similar 
during both observations and correspond to an average value for the cts/frametime/node
of $\sim 0.04$ and we do not worry about pileup.
The calibration with respect to charge transfer inefficiency (CTI) in
timed exposure (TE) mode cannot be used for CC mode, as here the short clocking
time changes the charge trap pattern in the device slightly reducing the effects of CTI.
Although the fast clocking limits CTI, 
we must manually adjust the order sorting tables to capture trailing charges.
Additionally, the continuous clocking morphs some of the flight event grades into flight grades 
outside the telemetered standard
grade system. The result is 
a continuous decrease in effective area towards higher energies. 
In contrast to the effects of pileup, however, discrete line features
remain preserved. 

For the purpose of this letter we actually do not need the most precise area calibration as we
are not fitting physical continuum models.
To fit emission lines we used the interactive analysis software ISIS
\footnote{see \url{http://space.mit.edu/ASC/ISIS/}}~\citep{houck2000}, 
applied standard wavelength 
redistribution matrix files (RMF),
and generated standard ancillary response files (ARFs) 
\footnote{see \url{http://asc.harvard.edu/ciao/threads/}}.
For all the observations we
generated spectra and analysis products for the 1st orders only.
We co-add +/-1st orders of both gratings, which results in a spectral grid of close to 0.021~\AA~
and an average of several $\sim 10^4$ cts bin$^{-1}$ between 1.8 and 14~\AA. For the spectral
analysis we always co-add both observations.
We determine an unabsorbed~\citep{juett2004} source flux in the
HETG spectra of $(1.53\pm0.25)\times10^{-8}$ \ergcm.

\section{Light Curves and Color-Color Diagrams
\label{cfluxes}}

The top panel in Figure~\ref{lightcurves} shows the light curve of obs. 1, the
one in obs. 2 is very similar.
The source 
was so strong that many frames were dropped during transmission. 
In order to properly deal with this effect we employed the
``aglc" package which is contributed software linked to from the CIAO pages
\footnote{see \url{http://asc.harvard.edu/ciao/download/scripts/}}. 
Frame drops vary between 10$\%$ and 70$\%$ in single CCDs at a time.

Both light curves appear morphologically very similar and feature a steady rise,
periods of steady high flux, and various rapid dipping periods. 
The HETG count rate varies between 200 and 300 cts s$^{-1}$
amounting to an equivalent of 0.47 Crab on average. 
Bins of 500 s thus result in very small statistical uncertainties.
Due to the limited bandpass in the HETGS we have use slightly
different wavelength bands to compute hardness ratios with respect
to studies done with \exo, \gin, and \xte.
We chose f$_{xs}$=flux(0.5 keV -- 2.5 keV),
f$_{xm}$=flux(2.5 keV -- 4.5 keV), and f$_{xh}$=flux(4.5 -- 8.0 keV) and thus
for the hard ratio f$_{xh}$/f$_{xm}$ and the soft ratio f$_{xm}$/f$_{xs}$. 

\input{tab2}


The color-color diagram corresponding to the light curve of obs. 1 is shown at the 
bottom of Figure~\ref{lightcurves}. In each of the two observations the entire Z-track is
covered once. The hardness variability
behavior follows the established pattern seen in Cyg X-2 in recent years with
HB and NB corresponding to flux rise and stability and the FB corresponding
to dipping. 

\section{The Properties of Bright Emission Lines
\label{hetganal}}

The unfolded flux spectrum in the range between 1.5 and 25~\AA~ 
exhibits a variety of line features. These consist of narrow absorption lines at
longer wavelengths and broad emission lines at shorter wavelength. 
The line absorption is attributed to the interstellar medium
will be dealt with in a separate paper~\citep{yao2008}.
In this letter we focus entirely on the 
measurement of a few of the isolated and brightest emission lines at short wavelengths ($<$ 14~\AA).  
These lines are well resolved with respect to the instrument resolution, which allows us
to determine excess fluxes independent of the underlying continuum and thus for the 
first time provide a direct diagnostic of plasma properties.

The lines are fit with single Gaussian functions and a local continuum generated within
$\pm$ 1.0~\AA~of the line center.  
Except for Fe, the identifications indicate Lyman $\alpha$ emissions from H-like
ions of \nex, \mgxii, \sixiv, and \sxvi. In the case of Fe there is \fexxv, \fexxvi, and possibly \fexxiv.
The \nex is blended into a very broad blend ranging from 10.5 to 12.5~\AA~ which
include also \fexxiv and lower Fe transitions, and here we cannot readily isolate
its line properties. The brightest lines are isolated we expect no significant blends.
The line centroids appear slighty blue-shifted with respect to the expected rest wavelengths. This
is anticipated due to \cygx2's systemic motion towards the Sun of $\sim$220 km s$^{-1}$~\citep{cowley1979}. 
Table~2 summarizes the properties of the brightest emission lines, with the line wavelengths
corrected for the systemic motion. The lines appear within less than $\sim 12$ mA of the rest
wavelengths on average. 
The lines from \mgxii, \sixiv, \sxvi
and the Fe K region are most significant. Figure~\ref{hetgspectra} shows
the spectrum plus residuals near Fe K as well as three examples of the line significances on a velocity
scale with respect to the rest wavelength. 
Doppler velocities deduced from the H/He-like lines range from 1100 at \fexxvi to 2730
\kmps at \mgxii. Also note that H-like line emissions generally contain two narrowly separated
line components and we 
acccounted for this separation in the calculation of the Doppler velocities. 
The \fexxv appears wider because it is a line triplet and the widths of its line 
components are consistent
with the other lines. The line width for \fexxv in Table~2 was obtained by fitting
a single function to the entire triplet. In 
Figure~\ref{hetgspectra} we fixed the relative wavelengths to the respective resonance,
intercombination, and forbidden line with line widths fixed to the value found
for \fexxvi. This shows that in 
\fexxv the bulk of the line
flux is associated with the intercombination line (strong green line). 

The spectrum harbors some narrower emission lines as well at wavelengths below 8~\AA~ with line widths
of several hundred km s$^{-1}$. Most significant is the one at 7.98~\AA\ which identifies
with the shortest wavelength of the \fexxiv (Li-like) ion sequence. The second strongest
is the one at 7.80~\AA\ which consistent with a strong intercombination
line of \alxii, with no detection of a resonance or forbidden line component.
The detection of the \alxii intercombination line as well as the non-detection
of the forbidden line provides a simple R ratio estimate. The R ratio is an important
density diagnostic and is defined as the ratio of forbidden to intercombination
line flux. Based on the continuum flux we can put an upper limit
to the forbidden line flux to 2.3$\times10^{-6}$ ph s$^{-1}$~cm$^{-2}$ leading to an
upper limit of the R ratio of 0.06 with 90$\%$ confidence of 0.01. 
In the case of Fe XXV we can use the single fit
components of the intercombination line (1.2$\times10^{-4}$ ph s$^{-1}$~cm$^{-2}$) and 
the forbidden line (0.3$\times10^{-4}$ ph s$^{-1}$~cm$^{-2}$) to estimate R $\sim$ 0.3,
well below unity.

\section{Line Fluxes along the Z-Pattern
\label{dlines}}

We also integrated the spectra along the three spectra branches. The time in the HB branch 
is about 15$\%$, in the NB 30$\%$, and in the FB is about 55$\%$ of the total exposure.  
The narrower lines at 7.98~\AA\ and 7.80~\AA\ were present in all branches with some
variability. The broad \nex line blend between 10.5 and 12.5~\AA\ is detected in all branches as well,
but appears most prominent in the FB.
Figure~\ref{lineflux} shows the evolution of line fluxes of the brightest lines
which are \mgxii, \sixiv, \sxvi, and \fexxv (see Table~2). \mgxii is hardly detected
in the HB and NB, \sxvi is very weak in the HB. The others are significant in all branches.
The flux at \fexxv increases by a factor 4, at \mgxii and \sxvi
by an order of magnitude from the HB to the FB. We emphasize that the observed variation
pattern cannot be a result of the CTI grade losses, which would have yielded the opposite trend.
Application of the grade loss rates would actually enhance this pattern. 

\includegraphics[angle=0,width=8.5cm]{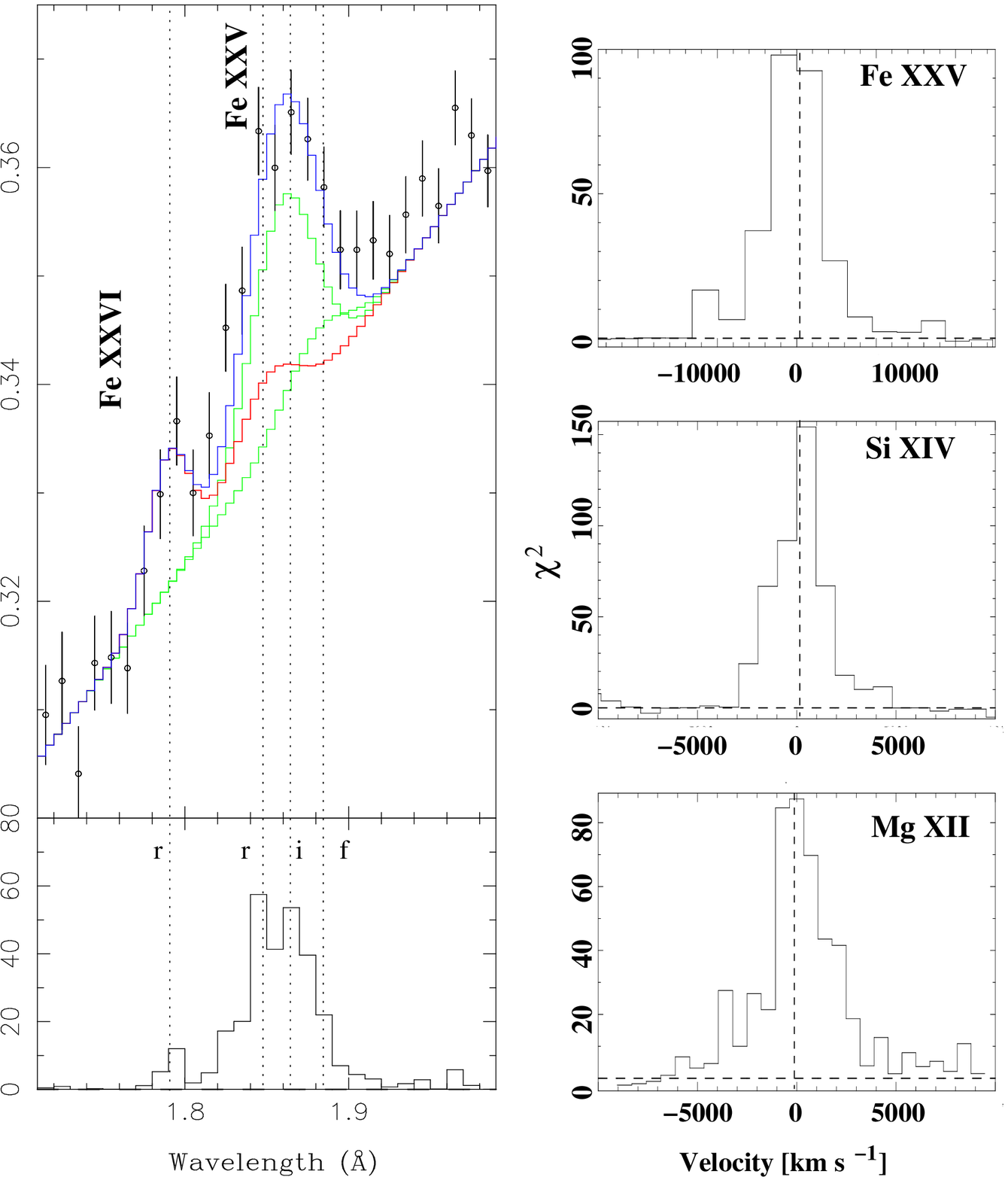}
\figcaption{{\bf Left:} Unfolded spectrum and residuals near the Fe K region. The blue line represents
the full line model, the red line represents the resonance lines,
the strong green line the intercombination,
and the weak green line the forbidden line of \fexxv.
{\bf Right:} Line residuals of \mgxii, \sixiv, and \fexxv
plotted on a velocity scale relative to the lines rest wavelengths.
\label{hetgspectra}} 
\vspace{0.2mm}

The current lack of calibration in cc-mode does not allow us to apply yet more detailed
models to determine the full ionization balance and its change from the HB to the NB.  
However by using the \it photemis \rm function within the photoionization code XSTAR
\footnote{see \url{http://heasarc.gsfc.nasa.gov/docs/software/xstar/xstar.html}}
to determine relative line strengths
we can set rough limits to the ionization parameter.
We find that the detected line emissivities require ionization parameters log $\xi$ of above 3.
For the variation from the HB to 
the FB we expect changes from 2.8 to 3.4, mostly based on the ratio of \fexxv to \fexxvi. 

\section{ADC Properties in \cygx2
\label{results}}

The preliminary analysis of a long exposure of the bright Z-source \cygx2 presented in this letter
produced a variety of new results. 
We detect a number of broad emission lines at short wavelengths corresponding to
\nex, \mgxii, \sixiv, \sxvi, as well as \fexxiv, \fexxv, and \fexxvi. The detections 
confirm the identifications from the \bbxrt , \asca , and \sax observations with respect to the
\fexxv and Fe L detections~\citep{smale1993, disalvo2002} and emissions from other abundant 
elements \citep{kuulkers1997}. For
the first time we can directly determine physical properties of the ADC as well
as true source luminosities along the Z-track. The 
emissivities indicate ionization parameters well above log $\xi$ = 3.
Such high ionization parameters imply plasma temperatures
in excess of 10$^6$ K.

\includegraphics[angle=0,width=8.5cm]{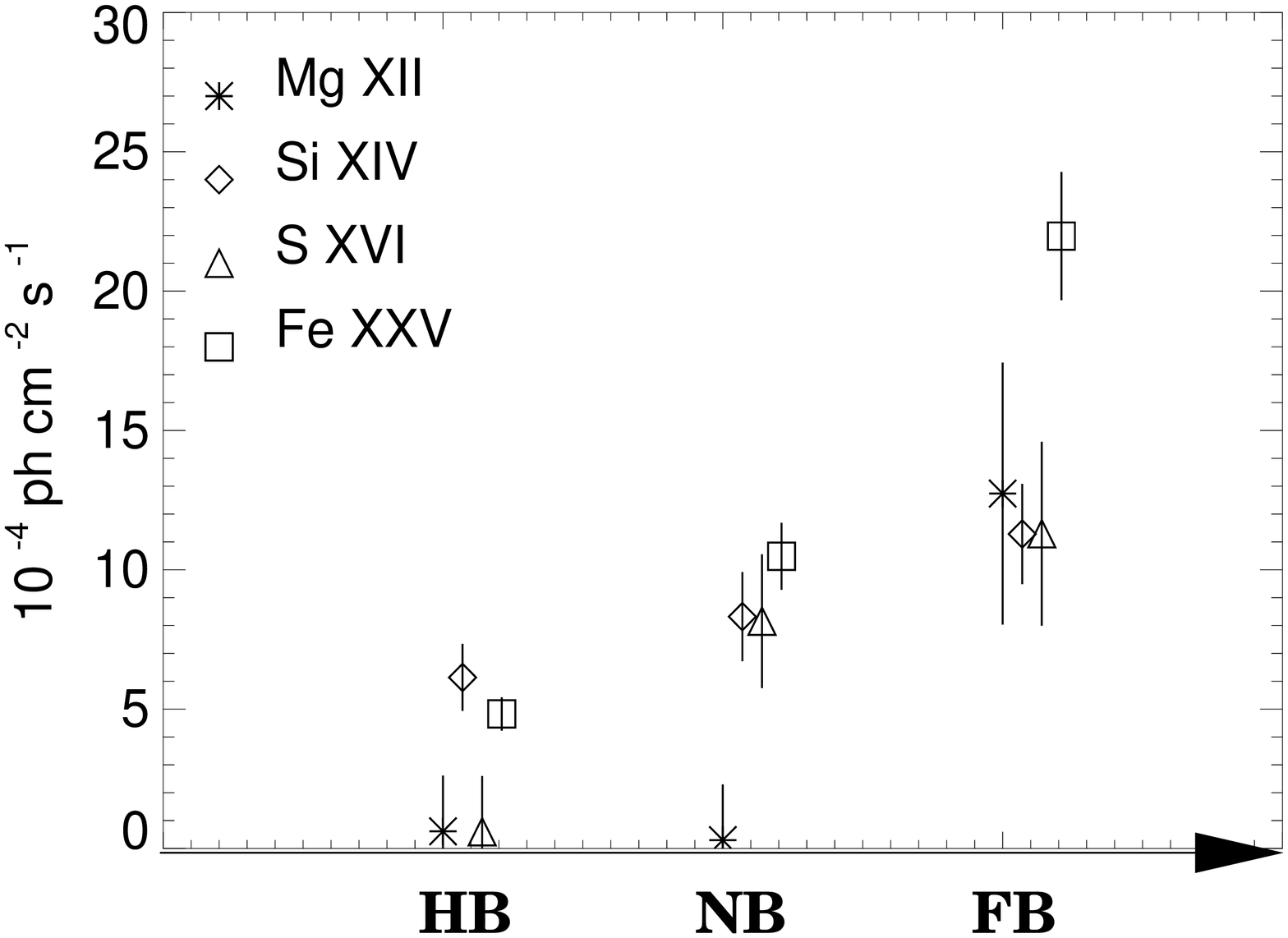}
\figcaption{Line fluxes of the brightest and broadest lines as
measured separately on the HB, NB, and FB.
\label{lineflux}}
\vspace{0.2mm}

We can use the line widths as well as the determined R ratios to determine the basic properties 
of the ADC in \cygx2. 
The lines are observed at rest and appear very broad with Doppler velocities 
between 1120 and 2730 \kmps, which at
an inclination of 60 degrees 
places the emissions at disk radii of 1.8 -- 11.0$\times 10^9$ cm from the neutron star.
These radii are consistent with recent model predictions for ADC sources \citep{mario2002}
and specifically for \cygx2 by \citet{church2006}. 
Furthermore, in equilibrium and at the high temperatures the 
He-like density diagnostics of photoionized plasmas 
behaves kinetically identical to the collisional case \citep{liedahl1999} and we can use
our R ratio and the density dependence in the Astrophysical Plasma Emission Database 
\footnote{see also \url{http://cxc.harvard.edu/atomdb/features$\_$density.html}} \citep{smith2001} to
determine a lower limit of 6.4($\pm1.4)\times10^{14}$ cm$^{-3}$ to the density of the corona. 
In the case of the \fexxv line APED does not provide the diagnostic, however its
estimated R ratio of less than unity indicates densities beyond 10$^{17}$ cm$^{-3}$, which
is the critical density for \fexxv~\citep{schulz2008}. These densities are too high to allow
for a direct view through this corona thus limiting the scale height above the accretion
disk to well below 30 deg for an assumed inclination of 60 deg. Stationary
ADCs with such scale heights have been modeled~\citep{mario2002} 
and observed for sources like \herx1, \4u1822, and \cirx1 
\citep{cottam2001, jimenez2005, schulz2008}.   

At a radius of $10^{10}$ cm, an average ionization parameter of 2000, and an average density
of $10^{15}$ cm$^{-3}$ 
we estimate a luminosity of 2$\times10^{38}$ \ergsec, consistent with
the average source luminosity during our observations of
1.8$\times10^{36} D^2$/[kpc] \ergsec at a source distance of 10.5 kpc. 

For the first time we find substantial line flux variability along the Z-pattern.
At coronal densities and radii as stated above, the
implied change in $\xi$ from $\sim 2.8$ to 3.4 provides us with an independent measure of the 
true change of the source luminosity along the Z-profile. 
The change
in average heating luminosity is then from 0.63$\times10^{38}$ \ergsec (0.35$\times L_{Edd}$) in the 
HB to 2.5$\times10^{38}$ \ergsec (1.40$\times L_{Edd}$) in the FB.
The direction of heating in such an ADC has immediate
implications with repect to recently published Z-models~\citep{church2008,lin2008}.
  
Future aspects of the analysis include a more detailed understanding of the various 
ionization regimes observed and their changes along the Z-track.
The next step in the analysis thus is to fit the spectra with physical models to obtain the full range
of ionization balances with specific attention to the complex line blend around 1 keV.

\acknowledgments
We thank all the members of the \chandra\ team for their enormous efforts.
We gratefully acknowledge the financial support of
Smithsonian Astrophysical Observatory contract SV1-61010 for the CXC. 


\end{document}

%% file: tab1.tex
\begin{center}
{\sc TABLE~1: OBSERVATIONS IN 2007}
\begin{tabular}{lccccc}
            & & & & & \\
\tableline
 Obsid &  Start Date & Start Time & Exp. & HETG 1st    &\\
       & [UT]        & [UT]       & [ks] & cts s$^{-1}$ &\\
\tableline
            & & & & & \\
 8170 & Aug 25 2007 & 17:45:20 &  70.2 & 272.23 &  \\
 8599 & Aug 23 2007 & 05:02:33 &  64.4 & 276.31 &  \\
            & & & & & \\
\tableline
\end{tabular}
\end{center}

%% file: tab2.tex
\begin{center}
{\sc TABLE~2: X-RAY LINE PROPERTIES} \\
\begin{tabular}{lccc}
 & & & \\
\hline
\hline
 ion & $\lambda_{\rm meas}$  & Flux$_{\rm line}$ & $v_{\rm D}$ \\
     &  \AA & (a) & km s$^{-1}$ \\
\hline
 & & & \\
Fe~XXVI           &   1.792$\pm$0.009 &  0.39$\pm$0.28 & 1120$\pm$870 \\
Fe~XXV            &   1.861$\pm$0.005 &  1.97$\pm$0.27 & 3450$\pm$710 \\
S~XVI~L$\alpha$   &   4.726$\pm$0.011 &  0.79$\pm$0.25 & 1860$\pm$1140 \\
Si~XIV~L$\alpha$  &   6.188$\pm$0.005 &  1.03$\pm$0.15 & 1610$\pm$290 \\
Al~XII           &   7.812$\pm$0.003 &  0.38$\pm$0.06 &  530$\pm$110 \\
Fe~XXIV          &   7.973$\pm$0.001 &  0.78$\pm$0.05 &  370$\pm$40 \\
Mg~XII~L$\alpha$  &   8.419$\pm$0.004 &  1.26$\pm$0.38 & 2730$\pm$480 \\
Ne~X~L$\alpha$    &   12.13           &  $\approxlt$1.3 &  $\approxlt$ 5600 \\
 & & & \\
\hline
\end{tabular}
\normalsize
\end{center}
(a) 10$^{-4}$ ph s$^{-1}$~cm$^{-2}$, uncertainties are 90$\%$ confidence\hfill